# ON THE PROBABILITY OF A CAUSAL INFERENCE IS ROBUST FOR INTERNAL VALIDITY

**Tenglong Li**

**Department of Biostatistics, Boston University**

**Email: litenglo@bu.edu**

**Kenneth A. Frank**

**Department of Counseling, Educational Psychology and Special Education, Michigan State University**

**Email: kenfrank@msu.edu**

**Abstract**

The internal validity of observational study is often subject to debate. In this study, we define the counterfactuals as the unobserved sample and intend to quantify its relationship with the null hypothesis statistical testing (NHST). We propose the probability of a causal inference is robust for internal validity, i.e., the PIV, as a robustness index of causal inference. Formally, the PIV is the probability of rejecting the null hypothesis again based on both the observed sample and the counterfactuals, provided the same null hypothesis has already been rejected based on the observed sample. Under either frequentist or Bayesian framework, one can bound the PIV of an inference based on his bounded belief about the counterfactuals, which is often needed when the unconfoundedness assumption is dubious. The PIV is equivalent to statistical power when the NHST is thought to be based on both the observed sample and the counterfactuals. We summarize the process of evaluating internal validity with the PIV into an eight-step procedure and illustrate it with an empirical example (i.e., Hong and Raudenbush (2005)).

Keywords: observational study, causal inference, internal validity, Bayesian statistics, sensitivity analysis

## 1-Introduction

Causal inferences are often made based on observational studies, which allow researchers to collect relatively large amounts of data with low cost per research question, compared to randomized experiments (Rosenbaum 2002; Shadish et al., 2002; Schneider et al. 2007). However, given observational studies do not employ randomization upon which causal inferences critically rely, their internal validity is often challenged and difficult to assess (Rosenbaum and Rubin, 1983b; Shadish et al. 2002; Rosenbaum, 2002, 2010; Imai et al. 2008; Murnane and Willett, 2011; Imbens and Rubin, 2015). In this paper, we inform debates about causal inferences from observational studies by quantifying the robustness of inferences from observational studies with regard to concerns about internal validity (Frank and Min 2007; Frank et al. 2013). We apply our approach to Hong & Raudenbush (2005) which estimated a negative effect of kindergarten retention on reading achievement. Although Hong and Raudenbush analyzed a nationally representative sample mitigating concerns about external validity, the treatments (i.e., retained in kindergarten versus promoted to the first grade) were not randomly assigned in this observational study, raising potential concerns about internal validity (Schafer and Kang, 2008; Allen et al. 2009; Hong, 2010; Frank et al., 2013).

To characterize concerns about internal validity in observational studies, we adopt the framework of Rubin Causal Model (RCM) (Holland, 1986; Rubin, 2008). A key concept of RCM is the potential outcome, which refers to the outcome of every subject under every possible treatment (Rubin 2007, 2008). A fundamental issue is that a subject can only choose one treatment at a time and thus only one potential outcome is observable. This renders all other potential outcomes missing (Rubin, 2005; Imbens and Rubin, 2015). In short, RCM recasts causal inference as a missing data problem where the missing outcomes are assumed to be missing at random (MAR) conditional on a set of covariates, an assumption known as "unconfoundedness" (Rosenbaum and Rubin, 1983a; Imbens, 2004).

Given the difficulty of justifying the unconfoundedness assumption, one may suspect the missing potential outcomes (i.e., counterfactual outcomes) are not MAR conditional on controlled covariates (Heckman, 2005; Rosenbaum and Rubin 1983b; Rosenbaum, 1987). As a result, the missing potential outcomes may not be comparable to the observed outcomes, which raises the following two questions: The first one is "what is your belief about the counterfactual outcomes" and the second is "what does your belief imply on the internal validity of your inference".

We leverage this logic to quantify the robustness of a causal inference based on one's belief about the counterfactual outcomes. To do this, we first define counterfactual outcomes as the unobserved sample and incorporate such unobserved sample into the observed sample to form the ideal sample, which, as indicated by its name, is ideal for making a causal inference (Sobel, 1996; Rubin, 2004, 2005; Frank et al. 2013). We further define the probability of a causal inference is robust for internal validity (henceforth, we abbreviate it as the PIV) based on the ideal sample as the robustness index of internal validity. Our analytical procedure aims to bound the PIV of an inference based on one's belief and inform the strength of internal validity based on such bound(s).

**2-Research setting**

This paper targets observation studies with two groups, i.e., the treatment group and the control group. Furthermore, we only consider observational studies with representative samples so that we can focus on internal validity. This paper focuses on the simple group-mean-difference estimator (referred to as the simple estimator henceforth) of an average treatment effect, which computes the difference between the adjusted mean treated outcome and the adjusted mean control outcome. The adjusted means can be calculated based on propensity score matching or stratification and perceived as valid estimators of true means of treated outcome and control outcome when the unconfoundedness assumption holds[1].

The PIV is rooted in null hypothesis statistical testing (NHST) context. To conduct a causal inference, the null hypothesis $H_0: \delta = \delta_0$ is assumed to be tested against the alternative hypothesis $H_a: \delta \neq \delta_0$. Here $\delta$ denotes the true average treatment effect and $\delta_0$ is often zero. Our framework can be easily modified for one-sided alternative hypothesis. Furthermore, the PIV is meaningful when the null hypothesis has been rejected based on the observed sample and we are interested in whether the null hypothesis would be rejected if the counterfactuals were known.

**3-Counterfactuals as the unobserved sample**

In this section, we will define the unobserved sample in terms of counterfactuals in RCM, in order to formalize our discussion of the PIV and its analysis.

**Definition 1**: **The unobserved sample** refers to the collection of the counterfactual outcomes of all sampled subjects. **The unobserved treated sample** refers to the collection of the counterfactual outcomes of the sampled subjects who actually received the control. **The unobserved control sample** refers to the collection of the counterfactual outcomes of the sampled subjects who actually received the treatment.

**Example:** The unobserved sample of Hong & Raudenbush (2005) is the collection of counterfactual reading scores of sampled students in their study. Specifically, this unobserved sample can be decomposed into the unobserved control sample which is the collection of reading scores of retained students had they all been promoted to first grade and the unobserved treated sample which is the collection of reading scores of promoted students had they all been retained in kindergarten.

---
[1] Using the propensity scores as controls in the model, to match cases, or to construct strata.

Figure 1 illustrates the conceptualization of the unobserved sample in Hong & Raudenbush (2005) for the simple estimator. The observed outcome $Y_{r,i}^{ob}$ symbolizes the reading score of any retained student whose counterfactual outcome is $Y_{p,i}^{un}$. Likewise, the observed outcome $Y_{p,j}^{ob}$ represents the reading score of any promoted student whose counterfactual outcome is $Y_{r,j}^{un}$. The unobserved sample consists of counterfactual outcomes $Y_{p,i}^{un}$ and $Y_{r,j}^{un}$.

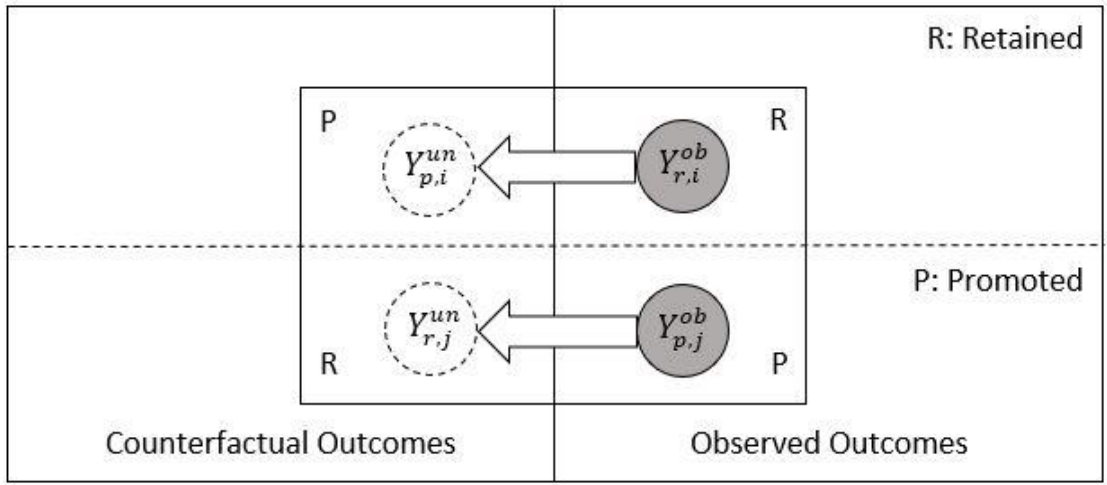

*Figure 1.* The unobserved sample in Hong & Raudenbush (2005) for the simple estimator

Finally, we define the ideal sample as follows:

**Definition 2**: **The ideal sample** refers to the combination of the observed sample and the unobserved sample. **The ideal treated sample** refers to the combination of the observed treated sample and the unobserved treated sample. **The ideal control sample** refers to the combination of the observed control sample and the unobserved control sample.

Drawing on the definitions above, we argue that it is the unobserved sample that induces the bias which undermines internal validity. The unobserved sample can be perceived as the gap between the observed sample and the ideal sample needed for insuring internal validity. The unconfoundedness assumption implies the unobserved sample is ignorable based on a set of covariates, i.e., the unobserved sample will essentially be the same as the observed sample conditional on the set of covariates. Given this assumption is frequently and constantly challenged, our goal is to quantify the robustness of the inference by discovering how the unobserved sample affect the NHST.

**4-The probability of a causal inference is robust for internal validity**

Frank et al. (2013) provided the following decision rules on whether a causal inference will be invalidated due to limited internal validity: Given a significant positive effect has been inferred in the observed sample, an inference will be invalidated if $\hat{\delta} > \delta^{\#} > \delta$. Given a significant negative effect has been inferred in the observed sample, an inference will be invalidated if $\hat{\delta} < \delta^{\#} < \delta$. Here $\hat{\delta}$ represents the estimated average treatment effect based on the observed sample and $\delta^{\#}$ is the threshold of rejecting the null hypothesis (and thus finding a significant effect). Since $\hat{\delta}$ is fixed and exceeds the threshold, the aforementioned decision rules can be simplified as $\delta < \delta^{\#}$ for a significant positive effect or $\delta > \delta^{\#}$ for a significant negative effect. The decision rules can be also interpreted in the opposite way: an inference cannot be invalidated if $\delta > \delta^{\#}$ for a significant positive effect or $\delta < \delta^{\#}$ for a significant negative effect. Drawing on this interpretation, the probability of a causal inference is robust for internal validity (PIV) is defined as the probability that an inference cannot be invalidated for the ideal sample $\mathbf{D^{id}}$. Specifically, the PIV is defined as follows for a significant positive effect:

$$P(\delta > \delta^{\#} | \mathbf{D^{id}}) \tag{1}$$

Likewise, the PIV is defined as follows for a significantly negative effect:

$$P(\delta < \delta^{\#} \mid \mathbf{D}^{id})  \qquad (2)$$

It's noteworthy that the PIV in (1) and (2) are actually the simplified version of

$P(\delta > \delta^{\#} \mid \hat{\delta} > \delta^{\#}, \mathbf{D}^{id})$ and $P(\delta < \delta^{\#} \mid \hat{\delta} < \delta^{\#}, \mathbf{D}^{id})$ respectively. Given the ideal sample must contain the observed sample, we can ignore the condition $\hat{\delta} > \delta^{\#}$ or $\hat{\delta} < \delta^{\#}$ as they should be conveyed by the ideal sample $\mathbf{D}^{id}$. The PIV essentially is the probability of rejecting the null hypothesis again for the ideal sample given the same null hypothesis has been rejected for the observed sample, when the counterfactual outcomes has been taken into consideration. This is tantamount to checking the impact of violation of the assumption of uncounfoundedness assumption on the PIV.

**5-Theoretical framework**

**5.1-The distribution of true average treatment effect δ**

**Theorem 1:** Assuming the treated outcome and the control outcome are independent and the variances of those two outcomes are given as $\sigma_t^2$ and $\sigma_c^2$ respectively, the distribution of δ based on the ideal sample would be:

$$\delta \mid \mathbf{D}^{id} \sim N(\theta_t - \theta_c, \phi_t + \phi_c) \qquad (3)$$

Where:

$$\theta_t = (1-\pi)\bar{Y}_t^{un} + \pi\bar{Y}_t^{ob}$$
$$\phi_t = \frac{\sigma_t^2}{n^{ob}}$$
$$\theta_c = \pi\bar{Y}_c^{un} + (1-\pi)\bar{Y}_c^{ob} \qquad (4)$$
$$\phi_c = \frac{\sigma_c^2}{n^{ob}}$$

Here we need to conceptualize the unobserved treated sample mean $\bar{Y}_t^{un}$ and the unobserved control sample mean $\bar{Y}_c^{un}$. For example, for Hong and Raudenbush (2005), $\bar{Y}_t^{un}$ is the mean reading score of the promoted students had they been retained in the kindergarten and $\bar{Y}_c^{un}$ is the mean reading score of the retained students had they been promoted to the first grade. The ideal treated and control sample means, i.e., $\theta_t$ and $\theta_c$, would then become the weighted average between the unobserved treated (or control) sample mean and the observed treated (or control) sample mean, while the weight is defined by the term $\pi$ which is the proportion of treated subjects in the observed sample. The variances of $\theta_t$ and $\theta_c$ are $\phi_t$ and $\phi_c$, which are given by the variances $\sigma_t^2, \sigma_c^2$ and the observed sample size $n^{ob}$.

It's remarkable that theorem 1 can be proved in a either frequentist fashion or a Bayesian fashion (see proof in appendix), and therefore it has both frequentist and Bayesian interpretations (Li, 2018). In frequentist world, the unobserved sample is part of the ideal sample so that $\bar{Y}_t^{un}$ and $\bar{Y}_c^{un}$ will shape the distribution of $\delta$ as well as the final inference that are built on the ideal sample. In Bayesian world, the prior is conceived to be built on the unobserved sample and the likelihood is built on the observed sample, which is consistent with the literature stating that prior can be treated as a function of the data of particular interest (Diaconis and Ylvisaker, 1979, 1985; Frank and Min, 2007; Hoff, 2009; Pearl and Mackenzie, 2018). Strictly speaking, $\bar{Y}_t^{un}$ and $\bar{Y}_c^{un}$ are the prior parameters in the Bayesian world rather than the sample statistics that are sufficient for the distribution of $\delta$ in the frequentist world.

## 5.2-The relationship between the PIV and the unobserved sample means

Theorem 1 shows that the distribution of δ conditional on **D**$^{id}$ is determined by $\bar{Y}_t^{un}, \bar{Y}_c^{un}$ based on the unobserved sample as well as by $\bar{Y}_t^{ob}, \bar{Y}_c^{ob}, n^{ob}$ based on the observed sample. It further indicates that the PIV will solely depend on $\bar{Y}_t^{un}$ and $\bar{Y}_c^{un}$ holding the observed sample and the variances $\sigma_t^2, \sigma_c^2$ fixed. We formalize this relationship in the following theorem:

**Theorem 2:** Given the distribution of δ in theorem 1, the probit link of the PIV is a function of $\bar{Y}_t^{un}$ and $\bar{Y}_c^{un}$, conditional on the observed sample statistics $\pi, n^{ob}, \bar{Y}_t^{ob}, \bar{Y}_c^{ob}$ as well as the values of $\sigma_t^2, \sigma_c^2, \delta^{\#}$. Specifically, for a significant positive effect, we have:

$$probit(PIV) = \frac{\sqrt{n^{ob}}}{\sqrt{\sigma_t^2 + \sigma_c^2}} \left[ (1-\pi)\bar{Y}_t^{un} - \pi\bar{Y}_c^{un} + (\bar{Y}_t^{ob} + \bar{Y}_c^{ob}) \cdot \pi - \bar{Y}_c^{ob} - \delta^{\#} \right] \quad (5)$$

For a significant negative effect, we have:

$$probit(PIV) = \frac{\sqrt{n^{ob}}}{\sqrt{\sigma_t^2 + \sigma_c^2}} \left[ \pi\bar{Y}_c^{un} - (1-\pi)\bar{Y}_t^{un} - (\bar{Y}_t^{ob} + \bar{Y}_c^{ob}) \cdot \pi + \bar{Y}_c^{ob} + \delta^{\#} \right] \quad (6)$$

It's possible to conduct univariate analysis regarding either $\bar{Y}_t^{un}$ or $\bar{Y}_c^{un}$ and bivariate analysis regarding both, based on theorem 2. The univariate analysis is easier to start with, although it requires an assumed value for either $\bar{Y}_t^{un}$ or $\bar{Y}_c^{un}$ and bound the PIV based on a bounded belief of the other one. For example, given that the mean reading score of the retained students had they been promoted to the first grade (i.e., $\bar{Y}_c^{un}$) equals 45.2 and the upper bound of the mean reading score of the promoted students had they been retained instead (i.e., $\bar{Y}_t^{un}$) is 45.78 (their observed mean reading score), the lower bound of the PIV of Hong & Raudenbush (2005) is found to be 0.77. The bivariate analysis is built on a bounded belief about both $\bar{Y}_t^{un}$ and $\bar{Y}_c^{un}$ so that one can bound the PIV. For example, the lower bound of the PIV of Hong & Raudenbush (2005) is found to be 0.73 if $\bar{Y}_t^{un} \leq 45.78$ and $\bar{Y}_c^{un} \geq 44.77$.

### 5.3- $\delta^{\#}$ as a statistical threshold

The decision threshold $\delta^{\#}$ is a key element of the PIV and its relationship with the unobserved sample means. It could be either a fixed value that is pragmatically set based on transaction cost, policy implication or literature review or a statistical threshold that is a product between the critical value and the standard error. This section serves as a guide of the computation of $\delta^{\#}$ as a statistical threshold based on the ideal sample.

In general, when $\delta^{\#}$ is a statistical threshold, it equals $\pm 1.96 * se_{\hat{\delta}^{id}}$ aligned with the level of significance as 0.05. A prerequisite of computing $\delta^{\#}$ is determining $se_{\hat{\delta}^{id}}$, which refers to the standard error of the simple estimator of average treatment effect based on an ideal sample. $se_{\hat{\delta}^{id}}$ can be computed as follows:

$$se_{\hat{\delta}^{id}} = \sqrt{\phi_t + \phi_c} = \sqrt{\frac{\sigma_t^2 + \sigma_c^2}{n^{ob}}} \quad (7)$$

Resultantly, the probit functions in (5) becomes:

$$probit(PIV) = \frac{\sqrt{n^{ob}}}{\sqrt{\sigma_t^2 + \sigma_c^2}} \left[ (1-\pi)\bar{Y}_t^{un} - \pi\bar{Y}_c^{un} + (\bar{Y}_t^{ob} + \bar{Y}_c^{ob}) \cdot \pi - \bar{Y}_c^{ob} \right] - 1.96 \quad (8)$$

Likewise, the probit function in (6) becomes:

$$probit(PIV) = \frac{\sqrt{n^{ob}}}{\sqrt{\sigma_t^2 + \sigma_c^2}} \left[ \pi\bar{Y}_c^{un} - (1-\pi)\bar{Y}_t^{un} - (\bar{Y}_t^{ob} + \bar{Y}_c^{ob}) \cdot \pi + \bar{Y}_c^{ob} \right] - 1.96 \quad (9)$$

## 6-Example: The effect of kindergarten retention on reading achievement

### 6.1-Overview

Alexander et al. (2003) established kindergarten retention as a widespread phenomenon in the US and with profound impacts for both promoted children and retained children, and therefore it has long been a controversial issue. To address such controversy, Hong and Raudenbush (2005) conducted an analysis which combined a multilevel model controlling for logits of propensity scores and propensity score strata to evaluate the effects of kindergarten retention policy and actual kindergarten retention on students' academic achievement. They used a nationally representative sample which contained about 7639 students and 1070 schools. Drawing on this design, Hong and Raudenbush (2005) estimated the effect of kindergarten retention on students' reading achievement as -9.01 with standard error of 0.68, which amounted to a significant effect whose size is about 0.67. In light of this considerable effect, Hong and Raudenbush (2005) concluded that "children who were retained would have learned more had they been promoted" and therefore "kindergarten retention treatment leaves most retainees even further behind".

Nevertheless, the internal validity of Hong and Raudenbush (2005) is subject to debate because propensity score analysis is built on the assumption of unconfoundedness, which implies all confounding variables are able to be observed and controlled in the causal model. However, as argued by Frank et al. (2013), some confounding variables may not be fully measured and controlled, incurring selection bias in the estimate. In cases such that an omitted variable was negatively correlated with kindergarten retention and positively correlated with reading achievement, the negative effect of kindergarten retention could be biased, and thus their inference would be invalidated if such a variable were taken into account.

To address the concern about the internal validity of Hong and Raudenbush's inference, we propose an analytical procedure that employs the PIV and its relationship with counterfactuals. Specifically, this analytical procedure comprises eight steps: 1-specify the parameter values, 2-choose the decision threshold, 3-obtain the probit model, 4-bound either $\bar{Y}_t^{un}$ or $\bar{Y}_c^{un}$, 5-choose the threshold of the PIV for strong internal validity, 6-bound the PIV, 7-determine the strength of internal validity, 8-the bivariate analysis. The flowchart of this analytical procedure is displayed below:

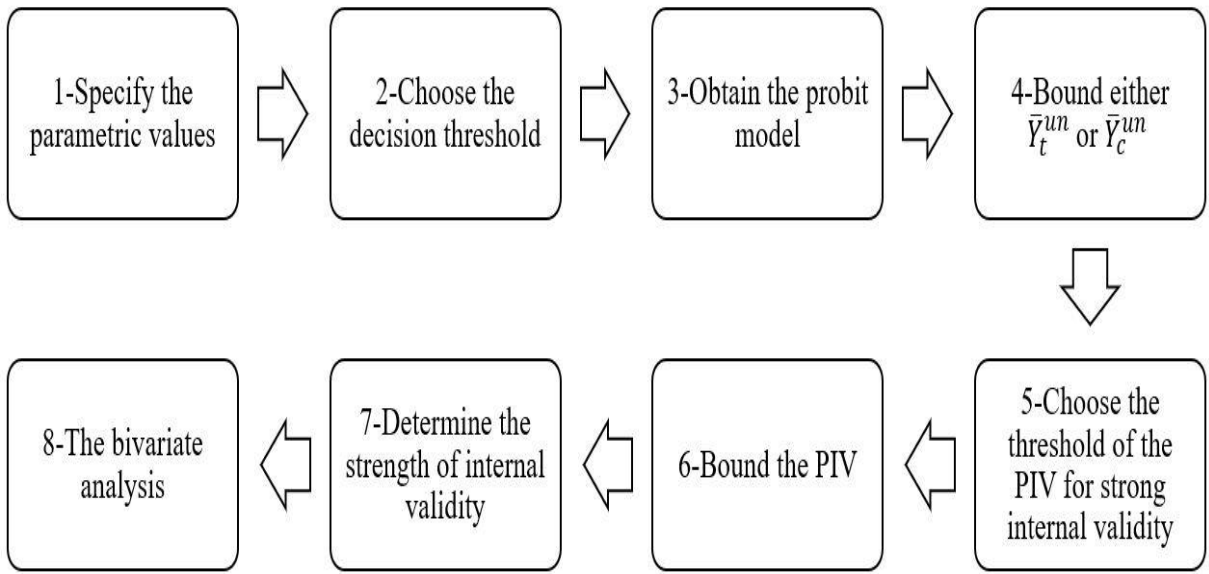

*Figure 2*: Flowchart of the analytical procedure for quantifying the robustness of an inference through the PIV

## 6.2-Quantifying the robustness of the inference of Hong & Raudenbush (2005)

1-Specify the parametric values: One will need to specify the values of $\sigma_t^2, \sigma_c^2$ as well as the observed sample statistics $\bar{Y}_t^{ob}, \bar{Y}_c^{ob}, n^{ob}, \pi$. The parameter values are specified as follows:

$\bar{Y}_t^{ob} = 36.77, \bar{Y}_c^{ob} = 45.78, \sigma_t^2 = 143.26, \sigma_c^2 = 138.83, n^{ob} = 7639, \pi = 0.0617$ (Frank et al. (2013)).

2-Choose the decision threshold: The threshold $\delta^{\#}$ is needed for deciding when the null hypothesis: $\delta = 0$ should be rejected. In this example we choose $\delta^{\#}$ to represent the statistical significance: $\delta^{\#} = -1.96 * se_{\hat{\delta}^{id}}$.

3-Obtain the probit model: Once the parameter values are plugged into the probit model (9), the probit model for Hong & Raudenbush can be explicitly written as:

$$probit(PIV) = 0.32\bar{Y}_c^{un} - 4.883\bar{Y}_t^{un} + 209.77 \qquad (10)$$

4-Bound either $\bar{Y}_t^{un}$ or $\bar{Y}_c^{un}$: This step asks one to state and bound his belief about one of the two mean counterfactual outcomes and make an assumption about the other. Given the inference of Hong & Raudenbush (2005) mostly informed $\bar{Y}_c^{un}$, i.e., the mean counterfactual reading score of the retained students, we decide to bound $\bar{Y}_t^{un}$ and assume $\bar{Y}_c^{un} = 45.2$. We choose this value because it is the grand sample mean so that $\bar{Y}_t^{un} - \bar{Y}_c^{un}$ measures the degree to which the counterfactual reading scores deviate from the null hypothesis: $\delta = 0$. The probit model (10) is thus simplified as follows:

$$probit(PIV) = 224.28 - 4.883\bar{Y}_t^{un} \qquad (11)$$

In this case, one need to ask himself "what could the mean reading score of the promoted students had they been retained instead (i.e., $\bar{Y}_t^{un}$) possibly be" when the mean reading score of the retained students had they been promoted instead (i.e., $\bar{Y}_c^{un}$) is assumed to be 45.2. It might be illuminating to reflect on the counterfactual outcomes based on the belief about the average retention effects for the retained students and for the promoted students, identified by $\bar{Y}_t^{ob} - \bar{Y}_c^{un}$ and $\bar{Y}_t^{un} - \bar{Y}_c^{ob}$ respectively. For example, given the average retention effect for the retained students is strongly negative (36.77-45.2 = -8.43), it is reasonable to think the average retention effect for the promoted students should be at least smaller than 0, as supported by literature in recent years. This leads to the upper bound for $\bar{Y}_t^{un}$ as 45.78.

5-Choose the threshold of the PIV for strong internal validity: We need to choose a cut-off value of the PIV such that internal validity is deemed strong whenever the PIV exceeds this cut-off value. It will be shown later that the PIV can be interpreted as the statistical power of retesting the null hypothesis: $\delta = 0$ for the ideal sample. Therefore, we recommend using PIV = 0.8 as the cut-off value, as it is often the cut-off value for good statistical power (Cohen 1988, 1992).

6-Bound the PIV: Given the upper bound for $\bar{Y}_t^{un}$ as 45.78, we derive the lower bound for the PIV as 0.77, based on the probit model (11). This means, given our belief that the mean reading score of the retained students had they been promoted instead is 45.2 and the mean reading score of the promoted students had they been retained instead is at most 45.78, the chance that Hong and Raudenbush's inference is robust for internal validity is at least 77%.

7-Determine the strength of internal validity: Given PIV = 0.8 as the threshold for strong internal validity, one would conclude that the internal validity of Hong & Randenbush's inference is strong enough since the PIV has a lower bound that quite close to 0.8 in this case.

8-The bivariate analysis: The bivariate analysis regarding both $\bar{Y}_t^{un}$ and $\bar{Y}_c^{un}$ is based on the belief that the average retention effects for the promoted students and for the retained students were both negative. The analysis also assumes the average retention effect for the retained students, which was originally estimated as -9 by Hong & Raudenbush, was overestimated. Therefore, the plausible region is defined based on the bounded beliefs that $\bar{Y}_t^{un} \leq 45.78$ and $36.77 \leq \bar{Y}_c^{un} \leq 45.78$. Figure 3 is used to illustrate our results. There are two key observations in figure 3: First, the lower bound of the PIV will be always higher than 0.8 as long as $\bar{Y}_t^{un} \leq 45.2$ and $\bar{Y}_c^{un} \geq \bar{Y}_t^{ob}$, i.e., to believe kindergarten retention has a negative impact on the retained students and the mean reading score of the promoted students had they been retained is lower than 45.2. This indicates the inference of Hong & Raudenbush is robust as long as kindergarten retention is believed to have considerably negative impact on the promoted students. Second, even if the kindergarten retention has minimal negative impact on the promoted students, the inference of Hong and Raudenbush (2005) would still be robust for internal validity as long as the average retention effect for the retained students was just slightly overestimated. For example, the lower bound of the PIV is 0.73 when the average retention effect for the retained students was believed to be at least -8 ($\bar{Y}_c^{un} \geq 44.77$), which is one point smaller in size than the original estimate. However, it would be risky to claim that the internal validity of Hong and Raudenbush (2005) is strong if $\bar{Y}_c^{un} \leq 44$ since the lower bound of the PIV in this case would drop below 0.64.

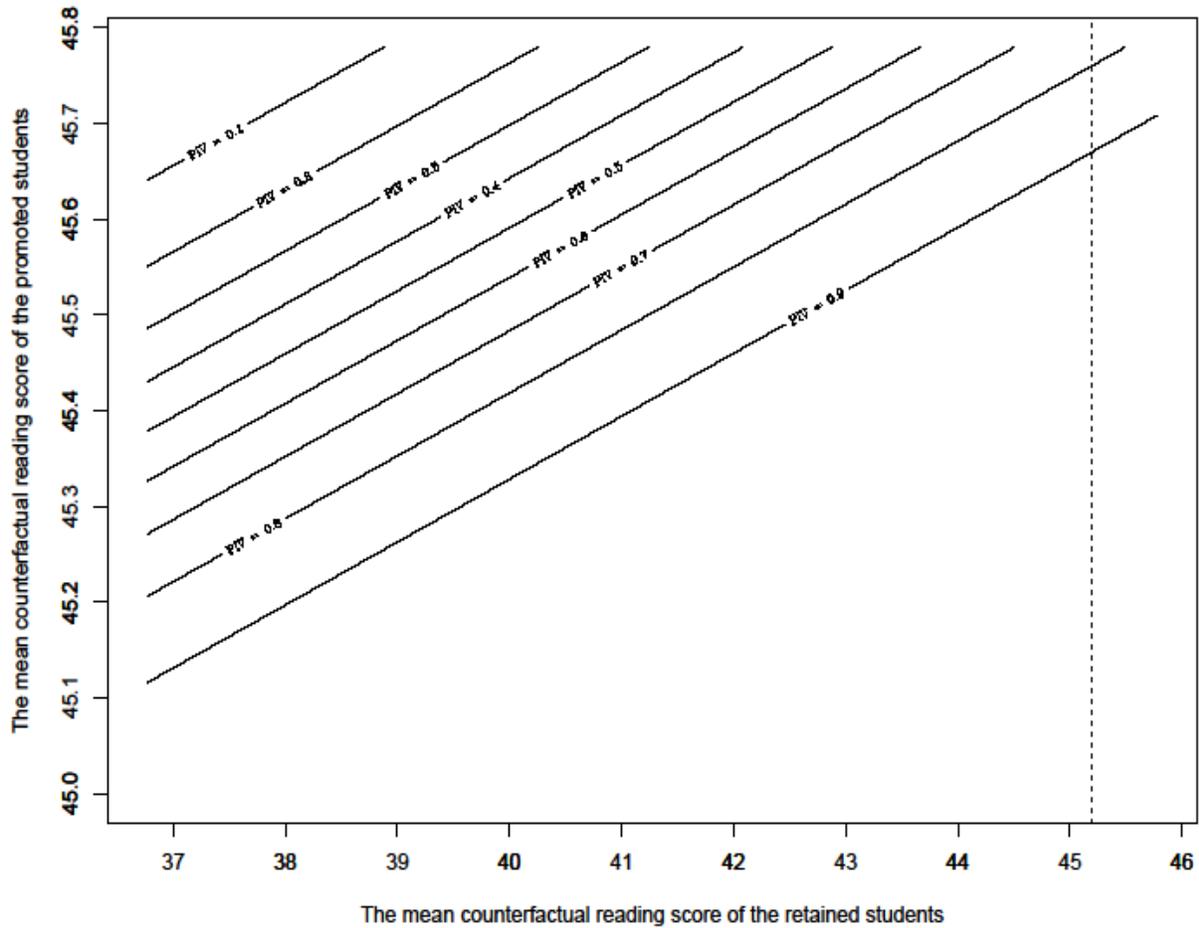

*Figure 3:* The contour plot of the PIV in the plausible region. The plausible region is defined based on the belief that the average retention effect for the promoted students should not be positive and the average retention effect for the retained students was overestimated, which means both $\bar{Y}_t^{un}$ and $\bar{Y}_c^{un}$ are smaller than 45.78. The vertical dashed line corresponds to our previous univariate analysis where $\bar{Y}_c^{un} = 45.2$.

### 6.3-Interpreting the PIV

In this section, we will explain how the PIV can be interpreted as the statistical power of retesting the null hypothesis: $\delta = 0$ had the counterfactual outcomes became observable. Our logic is as follows: Suppose one has rejected the null hypothesis in the observed sample, and concern about internal validity compels one to wonder if the null hypothesis would be rejected if the counterfactual outcomes were available. To conceptualize the above scenario, we would retest the null hypothesis based on the ideal sample and the original inference would be invalidated if we fail to reject the null hypothesis since this contradicts the significant positive/negative result found in the observed sample.

To unfold the general relationship between the PIV and retesting the null hypothesis, we provide table 1 which tabulates the thresholds of $\bar{Y}_t^{un}$ and $\hat{\delta}^{id}$ (estimate of average retention effect based on the ideal sample) when PIV equals 0.1, 0.2, …, 0.9 and $\bar{Y}_c^{un} = 45.2$. According to table 1, as the internal validity of Hong and Raudenbush (2005) becomes more robust (indicated by a larger PIV), the thresholds of $\bar{Y}_t^{un}$ should get smaller. This implies the average reading scores of the promoted students had they been retained as well as the estimate of the average retention effect on reading achievement in the ideal sample should be decreasing while the PIV is increasing. For example, the chance that Hong and Raudenbush's inference is robust for internal validity would be 80% or higher if $\bar{Y}_t^{un}$ is believed to be smaller than 45.76, which means $\hat{\delta}^{id}$ has to be smaller than -0.54. In order to make the chance that Hong and Raudenbush's inference is robust for internal validity even higher than 90%, one has to further believe $\bar{Y}_t^{un}$ is smaller than 45.67 and $\hat{\delta}^{id}$ is smaller than -0.62.

| Values of the PIV | Thresholds of $\bar{Y}_t^{un}$ | Thresholds of $\hat{\delta}^{id}$ |
|---|---|---|
| 0.1 | 46.19 | -0.13 |
| 0.2 | 46.1 | -0.22 |
| 0.3 | 46.04 | -0.28 |
| 0.4 | 45.99 | -0.33 |
| 0.5 | 45.93 | -0.38 |
| 0.6 | 45.88 | -0.43 |
| 0.7 | 45.82 | -0.48 |
| 0.8 | 45.76 | -0.54 |
| 0.9 | 45.67 | -0.62 |

*Table 1*: Thresholds of $\bar{Y}_t^{un}$ and $\hat{\delta}^{id}$ corresponding to PIV = 0.1, 0.2, …, 0.9 for the inference of Hong and Raudenbush, assuming $\bar{Y}_c^{un} = 45.2$.

Formally, the algebraic relationship between the PIV and retesting the null hypothesis based on the ideal sample can be written as follows:

When a significant positive effect has been concluded and $\delta^{\#} = 1.96 * se_{\hat{\delta}^{id}}$, we have:

$$probit(PIV) = T - 1.96 \tag{11}$$

When a significant negative effect has been concluded and $\delta^{\#} = -1.96 * se_{\hat{\delta}^{id}}$, we have:

$$probit(PIV) = -T - 1.96 \qquad (12)$$

T denotes the t-ratio corresponding to testing the null hypothesis: $\delta = 0$ versus the alternative hypothesis: $\delta \neq 0$ based on the ideal sample, i.e., $T = \dfrac{\hat{\delta}^{id}}{se_{\hat{\delta}^{id}}}$. From above formula, we know the only determinant of the PIV is T and that $\hat{\delta}^{id}$ is supposed to be negatively correlated with the PIV in Hong and Raudenbush, as manifested by table 1.

By definition, the PIV equals the statistical power of testing the null hypothesis: $\delta = 0$ versus the alternative hypothesis: $\delta = \hat{\delta}^{id}$ ($\hat{\delta}^{id} \neq 0$), which is illustrated by figure 4. It is clear that, as $\bar{Y}_t^{un}$ decreases, the estimate of average treatment effect in the ideal sample will be more extremely negative and resultantly the two distributions will be further apart conditional on the fixed parameter values and the statistical threshold. The PIV will then grow larger as those two distributions overlap less. Figure 4 vividly demonstrates how the PIV is equivalent to the statistical power when retesting the null hypothesis as if the counterfactual outcomes were available. Figure 4 informs that assessing internal validity through PIV can be linked to power analysis in the ideal sample.

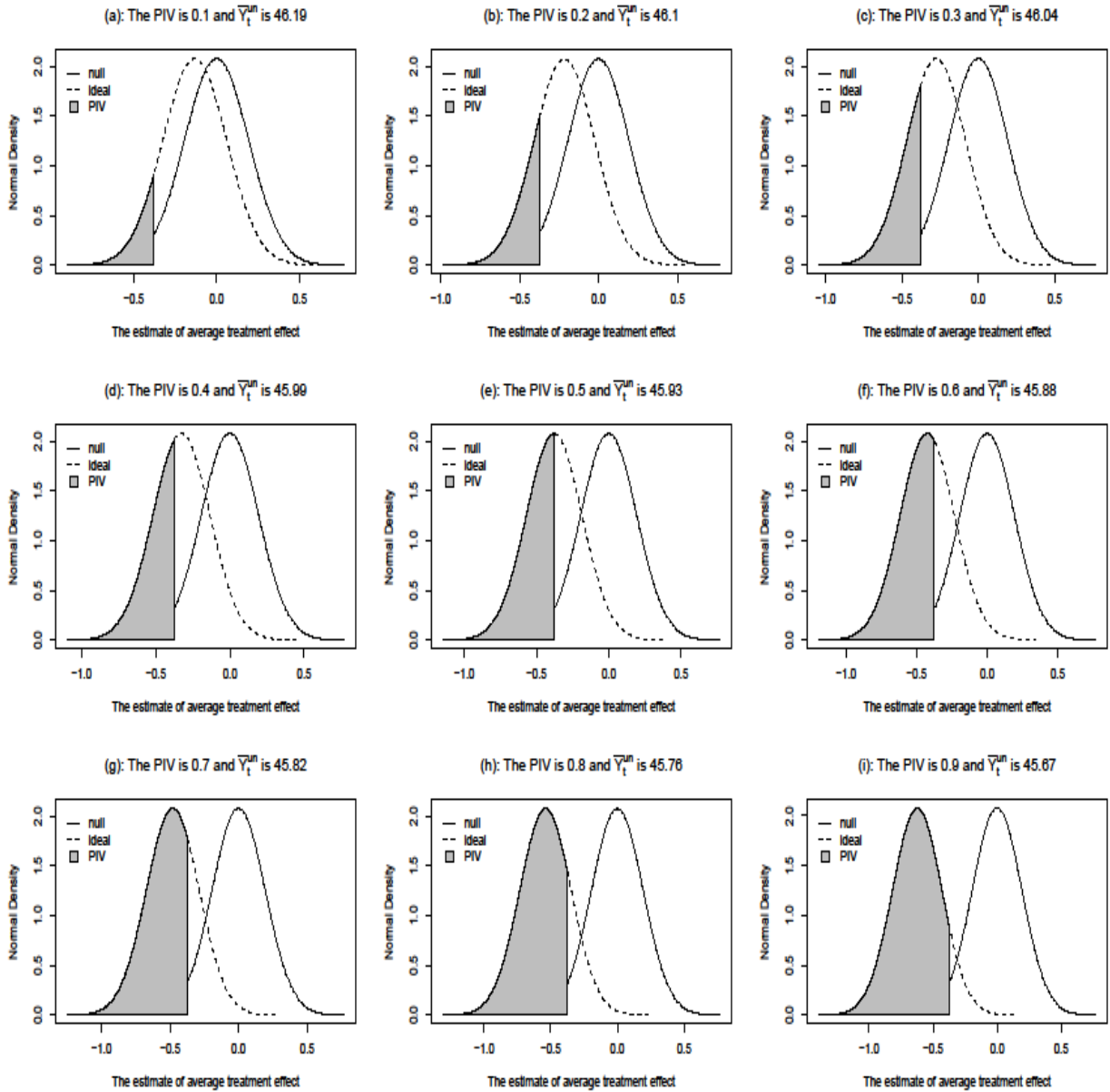

*Figure 4*: The relationship between the PIV and retesting hypothesis in the ideal sample for Hong and Raudenbush (2005), assuming $\bar{Y}_c^{un} = 45.2$. The solid curve represents the null hypothesis: $\delta = 0$ and the dashed curve represents the alternative hypothesis: $\delta = \hat{\delta}^{id}$. The grey shaded area symbolizes the PIV of Hong & Raudenbush.

# 7-Discussion

## 7.1-Literature review

Literature on sensitivity analysis: Sensitivity analysis (Rosenbaum and Rubin, 1983b; Rosenbaum, 1986, 1987, 1991, 2002, 2010) addresses the impact of an unobserved confounder on the estimates and inference for regression and nonparametric tests, and more importantly it connects the violation of unconfoundedness assumption to the violation of random assignment in matched pairs. Therefore, it informs the internal validity of a matching design. Other literature on sensitivity analysis has similar orientation towards missing confounders (Copas and Li, 1997; Lin et al., 1998; Robins et al., 2000; VanderWeele, 2008; Hosman et al., 2010; Masten and Poirier, 2018). The PIV shares the objective of checking the sensitivity of results to potential violation of the unconfoundedness assumption with the sensitivity analysis, but the PIV is not limited to a single type of design (like matching) or estimation (like regression). In fact, the PIV can be employed in any design that deemed appropriate in observational study where the counterfactuals is the main headache.

Literature on Bayesian sensitivity analysis: Bayesian sensitivity analysis (BSA) (McCandless et al. 2007, 2012; McCandless and Gustafson, 2017) parameterizes the models for explaining the outcome and the unmeasured confounder carefully so that it can identify the key parameters of confounding effect and examine their impacts on the estimate of treatment effect under a Bayesian framework. BSA has two main advantages: First, the data augmentation in Bayesian modeling allows one to build a model for the unobserved confounder and repeatedly draw random samples of it. As a result, one would get expected distributions of the confounding and treatment effect parameters. Additionally, BSA offers modeling flexibility through prior specification. Comparing to BSA, the implementation and interpretation of the analysis for the PIV would be much easier as BSA is built on complicated MCMC algorithm.

Literature on the robustness indices of causal inferences: The robustness indices of causal inferences (Frank, 2000; Frank et al., 2013) quantify the strength of internal validity in terms of the impact of an unmeasured confounding variable or the proportion of observed cases can be replaced by the null cases that an inference can afford. The PIV is inherently connected to both papers as it starts with the decision rules and the missing data perspective shared by Frank et al. (2013) and relies on the relationship between the estimate of average treatment effect and the NHST, which has been studied by Frank (2000). The PIV is different from the robustness indices because it requires a bounded belief about the counterfactual outcomes and it is a probabilistic index which is shown to be equivalent to the statistical power.

Literature on bounding treatment effect: Bounding treatment effect is proposed by acknowledging the issue of non-identification of the estimate of average treatment due to counterfactual outcomes. (Manski, 1990, 1995; Manski and Nagin, 1998). Different bounds of treatment effect can be obtained by imposing different assumptions on the counterfactuals, and the bounds of treatment effect would be tightened by making stronger assumption(s). Both the PIV and the bounds of treatment effect proposed by Manski consider the situations when the unconfoundedness assumption is implausible so that one has to form a belief about counterfactual outcomes. The key difference between the two approaches is the bounds of treatment effect does not directly address the probability that an inference would be still valid based on a set of assumption(s), as the PIV does. The bounded belief about counterfactual outcomes, which is the input for the PIV, is typically generated along with the bounds of treatment effect under the framework developed by Manski.

Literature on replication probability: Various replication probabilities have been proposed for two main reasons: First, they purpose safeguarding readers from the misguidance and misinterpretation of p-values. Second, they are used to accentuate that the true scientific significance is about replicability rather than statistical significance (Greenwald et al, 1996; Posavac, 2002; Shao and Chow, 2002; Killeen, 2005; Boos and Stefanski, 2011). The PIV is in fact the probability of replicating a significant result in observational study, and it is more akin to $p_{rep}$ (Killeen, 2005; Iverson et al., 2010) which is the probability of obtaining an effect with the same sign as the observed one. Different from $p_{rep}$ and all other replication probabilities, the PIV takes counterfactual outcomes into consideration and therefore it is not a function of p-value. Therefore, it does not inherit any weakness from p-value like most proposed replication probabilities do (Doros and Geier, 2005).

## 7.2-Conclusion

Founded on Rubin Causal Model (RCM), we began by defining the unobserved sample as the collection of counterfactual outcomes and the ideal sample as the collection of all the potential outcomes of the observed sample. It's worth emphasizing that the ideal sample is sufficient for securing internal validity and based on the ideal sample the null hypothesis is thought to be tested against the alternative hypothesis. The probability of a causal inference is robust for internal validity, i.e., the PIV, is thus defined in this scenario as the probability of rejecting the same null hypothesis again in the ideal sample given it has been rejected in the observed sample. This study recasts the assessment of internal validity as the task of bounding the PIV of an inference based on a bounded belief about the counterfactual outcomes.

This paper makes three main contributions to the field: First, it prompts researchers to conceptualize the counterfactual outcomes and form bounded belief about them. This will foster critical thinking as well as scientific discourse about internal validity since people can use the PIV to understand under what circumstances and to what degree internal validity will be robust. Second, the PIV has an intuitive interpretation. It is the statistical power of testing the hypothesis $H_0: \delta = \delta_0$ versus $H_a: \delta = \hat{\delta}^{id}$ in the ideal sample. Therefore, the PIV is pragmatic as it informs how counterfactual outcomes (and thus internal validity) influence the validity of a decision. Third, the modeling framework for the PIV is simple enough for empirical researchers and has both frequentist and Bayesian flavors.

Future work should focus on extending this model in two aspects: First, future work should revise the current model for subpopulations which are either non-normal or heterogeneous in nature, as the normality assumption is unlikely to hold in this case. Second, built on the framework which informs how counterfactuals affect the NHST through the PIV, future work needs to delve deeper into why counterfactuals change, which may due to the omit of confounding, the violation of SUTVA or measurement error.

# Appendix

## Proofs of Theorem 1 and Theorem 2

Proof of theorem 1:

First, the distribution of $\delta$ could be derived based on the following pivotal quantity:

$$\frac{\bar{Y}_t^{id} - \bar{Y}_c^{id} - \delta}{\sigma_{\bar{Y}_t^{id} - \bar{Y}_c^{id}}} \sim N(0,1) \tag{A1}$$

The pivotal quantity (A1) is built on the central limit theorem with the belief that the simple estimator $\bar{Y}_t^{id} - \bar{Y}_c^{id}$ should be unbiased for the true average treatment effect $\delta$.

Given the unobserved treated sample size is $n_t^{un}$ and the unobserved treated sample mean is $\bar{Y}_t^{un}$, it is straightforward to write the ideal treated sample mean $\bar{Y}_t^{id}$ as below:

$$\bar{Y}_t^{id} = \frac{n_t^{un}}{n_t^{ob} + n_t^{un}} \bar{Y}_t^{un} + \frac{n_t^{ob}}{n_t^{ob} + n_t^{un}} \bar{Y}_t^{ob} \tag{A2}$$

Similarly, the ideal control sample mean $\bar{Y}_c^{id}$ is written as follows:

$$\bar{Y}_c^{id} = \frac{n_c^{un}}{n_c^{ob} + n_c^{un}} \bar{Y}_c^{un} + \frac{n_c^{ob}}{n_c^{ob} + n_c^{un}} \bar{Y}_c^{ob} \tag{A3}$$

The standard deviation associated with the simple estimator is derived as below:

$$\sigma_{\bar{Y}_t^{id} - \bar{Y}_c^{id}} = \sqrt{\frac{\sigma_t^2}{n_t^{ob} + n_t^{un}} + \frac{\sigma_c^2}{n_c^{ob} + n_c^{un}}} \tag{A4}$$

Furthermore, because $n_t^{un} = n_c^{ob}$ and $n_c^{un} = n_t^{ob}$ in observational studies, we can simply the expressions in (A2) through (A3) as follows:

$$\bar{Y}_t^{id} = (1-\pi)\bar{Y}_t^{un} + \pi \bar{Y}_t^{ob}$$
$$\bar{Y}_c^{id} = \pi \bar{Y}_c^{un} + (1-\pi)\bar{Y}_c^{ob}$$
$$\sigma_{\bar{Y}_t^{id} - \bar{Y}_c^{id}} = \sqrt{\frac{\sigma_t^2}{n^{ob}} + \frac{\sigma_c^2}{n^{ob}}}$$

(A5)

given $\pi = \dfrac{n_t^{ob}}{n^{ob}}$ and $n^{ob} = n_t^{ob} + n_c^{ob}$. Finally, we can just plug (A5) in (A1), which yields the distribution (3) whose mean and variance are defined by (4).

Theorem 1 can also be proved under a Bayesian framework. Assuming the prior and the likelihood for the treated outcome is as follows:

$$\mu_t \sim N(\bar{Y}_t^{un}, \frac{\sigma_t^2}{n_c^{ob}})$$
$$Y_t \sim N(\mu_t, \sigma_t^2)$$

(A6)

And the prior and the likelihood for the control outcome is as follows:

$$\mu_c \sim N(\bar{Y}_c^{un}, \frac{\sigma_c^2}{n_t^{ob}})$$
$$Y_c \sim N(\mu_c, \sigma_c^2)$$

(A7)

The posterior distribution for $\mu_t$ is as follows:

$$\mu_t \mid \mathbf{D^{ob}} \sim N((1-\pi)\bar{Y}_t^{un} + \pi\bar{Y}_t^{ob}, \frac{\sigma_t^2}{n^{ob}}) \tag{A8}$$

Furthermore, the posterior distribution for $\mu_c$ is as follows:

$$\mu_c \mid \mathbf{D^{ob}} \sim N(\pi\bar{Y}_c^{un} + (1-\pi)\bar{Y}_c^{ob}, \frac{\sigma_c^2}{n^{ob}}) \tag{A9}$$

Assuming the treated and the control outcomes are independent, the posterior distribution of $\delta$ is equivalent to $\mu_t \mid \mathbf{D^{ob}} - \mu_c \mid \mathbf{D^{ob}}$ and has the identical form as (3).

Proof of theorem 2:

When a significant positive effect has been concluded, the PIV can be expressed by $\alpha$, $\bar{Y}_c^{un}$ and $\pi$ as follows, drawing on theorem 1:

$$PIV = P(\delta > \delta^\# \mid \mathbf{D^{id}}) =$$

$$P\left( \frac{\delta - (\bar{Y}_t^{id} - \bar{Y}_c^{id})}{\sqrt{\frac{\sigma_t^2 + \sigma_c^2}{n^{ob}}}} > \frac{\delta^\# - (\bar{Y}_t^{id} - \bar{Y}_c^{id})}{\sqrt{\frac{\sigma_t^2 + \sigma_c^2}{n^{ob}}}} \;\Big|\; \mathbf{D^{id}} \right) = 1 - \Phi\left( \frac{\delta^\# - (\bar{Y}_t^{id} - \bar{Y}_c^{id})}{\sqrt{\frac{\sigma_t^2 + \sigma_c^2}{n^{ob}}}} \right)$$

$$= \Phi\left( \frac{[(1-\pi)\bar{Y}_t^{un} + \pi\bar{Y}_t^{ob} - (1-\pi)\bar{Y}_c^{ob} - \pi\bar{Y}_c^{un}] - \delta^\#}{\sqrt{\frac{\sigma_t^2 + \sigma_c^2}{n^{ob}}}} \right) \tag{A10}$$

$$= \Phi\left( \frac{\bar{Y}_c^{un}(1-\pi) \cdot \alpha + (\bar{Y}_t^{ob} + \bar{Y}_c^{ob} - \bar{Y}_c^{un}) \cdot \pi - \bar{Y}_c^{ob} - \delta^\#}{\sqrt{\frac{\sigma_t^2 + \sigma_c^2}{n^{ob}}}} \right)$$

From (A10), the probit model for PIV can be derived as identical to (5).

Likewise, when a significant negative effect has been concluded, the PIV is expressed as follows:

$$PIV = P(\delta < \delta^{\#} \mid \mathbf{D^{id}}) =$$

$$P\left(\frac{\delta - (\bar{Y}_t^{id} - \bar{Y}_c^{id})}{\sqrt{\frac{\sigma_t^2 + \sigma_c^2}{n^{ob}}}} < \frac{\delta^{\#} - (\bar{Y}_t^{id} - \bar{Y}_c^{id})}{\sqrt{\frac{\sigma_t^2 + \sigma_c^2}{n^{ob}}}} \mid \mathbf{D^{id}}\right) = \Phi\left(\frac{\delta^{\#} - (\bar{Y}_t^{id} - \bar{Y}_c^{id})}{\sqrt{\frac{\sigma_t^2 + \sigma_c^2}{n^{ob}}}}\right)$$

$$= \Phi\left(\frac{\delta^{\#} - [(1-\pi)\bar{Y}_t^{un} + \pi\bar{Y}_t^{ob} - (1-\pi)\bar{Y}_c^{ob} - \pi\bar{Y}_c^{un}]}{\sqrt{\frac{\sigma_t^2 + \sigma_c^2}{n^{ob}}}}\right) \quad (A11)$$

$$= \Phi\left(\frac{\bar{Y}_c^{un}(\pi - 1) \cdot \alpha - (\bar{Y}_t^{ob} + \bar{Y}_c^{ob} - \bar{Y}_c^{un}) \cdot \pi + (\bar{Y}_c^{ob} + \delta^{\#})}{\sqrt{\frac{\sigma_t^2 + \sigma_c^2}{n^{ob}}}}\right)$$

From (A11), the probit model for PIV can be derived as identical to (6).

33